\input harvmac
\sequentialequations
\def\vx{\vec{x}}
\def\vS{\vec{S}}
\def\vJ{\vec{J}}
\lref\manton{N.~Manton, ``The Force Between `t Hooft-Polyakov Monopoles,''
Nucl. Phys. B126 (1977) 525.}

\lref\mp{S.~D.~Majumdar, Phys. Rev. 72 (1947) 930; A. Papapetrou, Proc. Royal Irish Acad. A51
(1947) 191.} 

\lref\hartle{J.~B.~Hartle and S.~W.~Hawking, ``Solutions of the Einstein-Maxwell
Equations with Many Black Holes," Comm. Math. Phys. 26 (1972) 87.}

\lref\iwp{Z.~Perj\'es, ``Solutions of the Coupled Einstein 
Maxwell Equations 
Representing the Fields of Spinning Sources,'' Phys. Rev. Lett. 27 (1971) 1668.
\hfill\break
W.~Israel, and G.~A.~Wilson, ``A Class of Stationary Electromagnetic 
Vacuum Fields,'' 
J. Math. Phys. 13 (1972) 865.}

\lref\tod{K.~P.~Tod, ``All Metrics Admitting Supercovariantly Constant Spinors,''
Phys. Lett. 121B (1983) 241.} 

\lref\wald{R.~M.~Wald, ``Gravitational Spin Interaction,'' Phys. Rev. D6 (1972) 406.}

\lref\papapetrou{A.~Papapetrou, ``Spinning Test Particles in General Relativity I,'' Proc. Royal 
Society London A209 (1951) 248.}

\lref\dixon{W.~G.~Dixon, ``Dynamics of Extended Bodies in General Relativity III: Equations of
Motion,'' Proc. Royal Society (London) A314 (1970) 499.}

\lref\brill{D.~Brill, ``Electromagnetic Fields in a Homogeneous, Nonisotropic 
Universe,'' Phys. Rev 133
(1964) 845.}

\lref\whitt{B.~Whitt, ``Israel-Wilson Metrics,'' Ann. Phys. 161 (1985) 241.}

\lref\nutdual{M.~Demianski and E.~T.~Newman, Bull. Acad. Pol. Sci. 14 (1966) 653;
J.~S.~Dowker, ``The NUT Solution as a Gravitational Dyon'', Gen. Rel. Grav. 5 (1974) 603.}

\lref\zee{A.~Zee, ``Gravitomagnetic Pole and Mass Quantization,''
Phys. Rev. Lett. 55 (1985) 2379, Erratum Phys. Rev. Lett. 56 (1986) 1101.}

\lref\dowker{J.~S.~Dowker and J.~A.~Roche, ``The Gravitational Analogues of
Magnetic Monopoles,'' Proc. Phys. Soc. London 92 (1967) 1.}

\lref\rafael{R.~D.~Sorkin, ``Kaluza-Klein Monopole," Phys. Rev. Lett. 51 (1983) 87.}

\lref\gross{D.~J.~Gross and M.~J.~Perry, ``Magnetic Monopoles in Kaluza-Klein Theories,"
Nucl. Phys. B226 (1983) 29.}

\lref\paul{P.~K.~Townsend, ``The Eleven-Dimensional Supermembrane Revisited," Phys. Lett B350 (1995) 184.}

\lref\ae{P.~C.~Aichelburg, and F.~Embacher,
``Exact superpartners of N=2 supergravity solitons,'' Phys. Rev. D34 (1986) 3006.}

\lref\aeone{P.~C.~Aichelburg, and F.~Embacher, ``Supergravity Solitons I: General Framework,''
Phys. Rev. D37 (1988) 338.}

\lref\aetwo{P.~C.~Aichelburg, and F.~Embacher, ``Supergravity Solitons II: the Free Case,''
Phys. Rev. D37 (1988) 911.}

\lref\aethree{P.~C.~Aichelburg, and F.~Embacher, ``Supergravity Solitons III: the Background 
Problem,'' Phys.Rev. D37 (1988) 1436.}

\lref\aefour{P.~C.~Aichelburg, and F.~Embacher,
``Supergravity Solitons IV: Effective Soliton Interaction,''
Phys. Rev. D37 (1988) 2132.}


\lref\bktw{V.~Balusubramanian, D.~Kastor, J.~Traschen and K.~Z.~Win, 
``The Spin of the M2-Brane and Spin-Spin Interactions via Probe Techniques'', hep-th/9811037.}

\lref\horowitz{G.~T.~Horowitz and A.~Sen, ``Rotating Black Holes Which Saturate a Bogomolnyi
Bound,'' Phys. Rev. D53 (1996) 808.}

\lref\stelle{K.~S.~Stelle, ``BPS Branes in Supergravity'', 
in {\it Trieste 1997, High energy physics and cosmology}, hep-th/9803116} 

\Title{\vbox{\baselineskip12pt
\hbox{UMHEP-456}
\hbox{hep-th/9811088}}}
{\vbox{\centerline{\titlerm BPS Force Balances}
\bigskip
\centerline{\titlerm via Spin-Spin Interactions}
 }}
\centerline{
David~Kastor\foot{kastor@phast.umass.edu}
and Jennie~Traschen\foot{traschen@phast.umass.edu}  }
\bigskip
\medskip
\centerline{\it Department of Physics and Astronomy}
\centerline{\it University of Massachusetts}
\centerline{\it Amherst, MA 01003-4525 USA}
\bigskip
\centerline{\bf Abstract}
\medskip
We study two systems of BPS solitons in which spin-spin interactions are important in establishing
the force balances which allow static, multi-soliton solutions to exist.  Solitons in the
Israel-Wilson-Perjes (IWP) spacetimes each carry arbitrary, classical angular momenta. Solitons in the
Aichelburg-Embacher ``superpartner" spacetimes carry quantum mechanical spin, which originates in the zero-modes
of the gravitino field of $N=2$ supergravity in an extreme Reissner-Nordstrom background. In each case we find a
cancellation between gravitational spin-spin and magnetic dipole-dipole forces, in addition to the usual
one between Newtonian gravitational attraction and Coulombic electrostatic repulsion. In both cases, we
analyze the forces between two solitons by treating one of the solitons as a probe or test particle, with the
appropriate properties, moving in the background of the other. In the IWP case, the equation of motion
for a spinning test particle, originally due to Papapetrou, includes a coupling between the background curvature
and the spin of the test particle. In the superpartner case, the relevant equation of motion follows from a
$\kappa$-symmetric superparticle action. 
\medskip
\Date{November, 1998}
\vfill\eject
\newsec{Introduction}

The list of BPS solutions to the classical field equations of $N\ge 2$ supersymmetric 
field theories seems to grow daily.  One of the interesting features of
BPS solitons are the, sometimes quite intricate, force cancellations which make static 
multi-soliton solutions possible. Between BPS magnetic monopoles, for example,
magnetostatic repulsion is balanced by an attractive scalar Higgs force \manton .
Amongst extreme Reissner-Nordstrom black holes, the electrostatic repulsion is balanced 
by gravitational attraction in the well known Majumdar-Papapetrou (MP) multi-black hole spacetimes
\mp\hartle . In multi-intersecting brane spacetimes, the cancellation may involve a long list of 
static gauge and scalar forces as well as gravity (see {\it e.g.} \stelle).

It is interesting to know all the different types of interactions which may contribute to BPS force balances.
In this short paper, we will study the cancellation of forces in two systems of BPS solitons
and identify two additional interactions, gauge and gravitational spin-spin forces,
which contribute in each of these cases.
We will focus primarily on the Israel-Wilson-Perjes (IWP) spacetimes \iwp ,
which are multi-soliton solutions of $4$-dimensional Einstein-Maxwell theory. 
This theory may be regarded as the bosonic sector of $N=2$ supergravity and, in this context,
it has been shown that the IWP spacetimes preserve half the full spacetime supersymmetry \tod .

Like the MP spacetimes, which form a subclass of the IWP solutions, each IWP
soliton has charge equal to its mass, giving a balance of  electrostatic repulsion and
Newtonian gravitational attraction between pairs of solitons. 
Each IWP soliton, however, may also carry classical angular momentum of 
arbitrary magnitude and direction. 
For nonzero angular momentum, an IWP soliton also has a magnetic dipole
moment, and at large separation two IWP solitons are subject to the familiar
magnetic dipole-dipole force.  We will show that in the force balance between spinning
IWP solitons this magnetic dipole-dipole force is cancelled
by a gravitational spin-spin force. Our strategy is to consider the forces on a BPS
probe, or test particle, which has mass equal to its charge and also carries
arbitrary classical spin, moving in the background of an IWP spacetime. 
The equation of motion for such a charged, spinning test particle was derived in 
\papapetrou\dixon ,  essentially via taking the probe limit of a finite sized
object carrying  charge and angular momentum. 
 
We also present a second example in which the solitons carry angular momenta that are
quantum mechanical in nature - of order $\hbar$. The multi-soliton solutions in this case
are the ``superpartner'' spacetimes of Aichelburg and Embacher \ae\ constructed by acting with broken
supersymmetry  transformations on the MP solutions\foot{See reference \bktw\ for a recent similar treatment of
M2-brane superpartners.}. The angular momentum in this case has its origin in the zero modes of the gravitino
field in the MP background (see \bktw\ for discussion of this point).
The probe we consider to analyse the force balance in this case, following the series of papers
\aeone\aetwo\aethree\aefour , 
is a  $\kappa$-symmetric superparticle, which has charge equal to its mass,
carries intrinsic spin  and couples to the background  spacetime fields of $N=2$ supergravity. 
We will see that the force balance for the superparticle in the superpartner background
works out in just the same way it does for the classical BPS test
particle in the IWP background, 
with the spin of the classical probe being replaced by the quantum mechanical 
spin operator of the superparticle.


\newsec{The Israel-Wilson-Perjes Spacetimes}

The IWP spacetimes are stationary, solutions to the Einstein-Maxwell equations given by
\eqn\iwpspace{\eqalign{
&ds^2=-|f|^{-2}(dt+\omega_idx^i)^2+|f|^2\delta_{ij}dx^idx^j,\cr
&F_{ti}=\partial_i\Phi,\qquad F^{ij}=|f|^{-2}\epsilon^{ijk}\partial_k\xi,\qquad
\Phi+i\xi=1/f.\cr}}
Here, $f(\vec{x})$ is any complex solution to the flat $3$-dimensional Laplacian
\eqn\laplace{\nabla^2 f=0,}
and the $3$-dimensional vector $\vec\omega$ is a solution to
\eqn\offdiagonal{
\vec\nabla\times\vec\omega=i\left(f\vec\nabla f^{*}- f^{*}\vec\nabla f\right).}
Taking $f$ to be real and restricting to pointlike singularities, an otherwise arbitrary $f$ may be
written as \eqn\mpspace{f=1+\sum_{i=1}^N{M_i\over |\vec{x}-\vec{x}_i|}.}
These are the MP spacetimes\mp.  If $f$ has only one singularity, 
it is well known that the spacetime
is identical to the extreme Reissner-Nordstrom black hole, having mass equal to its charge.
For $N>1$, the real parameters $M_i$ and $\vec{x}_i$ correspond, loosely
speaking, to the masses and positions of a collection of extreme black holes,
each having charge $Q_i$ equal to its mass $M_i$. This identification is not
precise, in part,  because only the total mass of a spacetime is well defined.
However, especially in the limit that the objects are widely separated, the 
identification seems justified. Hartle and Hawking\hartle\ have shown that the
objects in the MP solutions have horizons and so are indeed black
holes.

As discussed in \hartle, there are two independent ways to make the metric function $f$ in \iwpspace\ 
complex. The parameters $M_i$ and/or the parameters $x_i$ in \mpspace\ may be taken to be
complex.  Taking $M_i=m_i+in_i$ and keeping the $x_i$ real, yields a collection
of solitons with masses $m_i$ and NUT charges $n_i$.  The condition of equal charge and 
mass for MP solitons is replaced in this case by the relation
\eqn\nutrelation{m_k+in_k=Q_k+iP_k,}
where $Q_k$ and $P_k$ are the electric and magnetic charges of the $k$th soliton. 
The case of a single IWP soliton with $M$ complex is identical to the 
extremal limit of the charged generalization of Taub-NUT space given by Brill
\brill.  

Keeping the mass parameters $M_i$ real, but allowing the components of the
vectors $\vx_i$ to be complex gives a collection of objects, each having
electric charge $Q_i$ equal to its mass $M_i$ as in the MP case. Additionally, 
each object now carries an angular momentum, whose magnitude and direction are
encoded in its complex position vector $\vec x_i$.  For example,  as shown in
the second reference in \iwp, taking a single object with complex position
vector $\vx_1=ia\hat{k}$ and real mass parameter $M$, gives a Kerr-Newman
spacetime with electric charge equal to its mass, and angular momentum vector
$\vec {J}=aM \hat{k}$ directed along the $z$-axis.  This is, of course, a naked
singularity rather than a black hole. It is shown in \hartle\ that this is
generically  the case when a position vector is
taken to be complex in \mpspace\foot{Note however that IWP-type solutions in higher dimensions are
known to have regular horizons \horowitz .}.

\newsec{IWP Force Balances}
We can think of an IWP solution as a collection of solitons, each of which is characterized
by a set of parameters; mass, electric and magnetic charges, NUT charge and angular momentum vector.
The positions of these objects stay fixed in time, so that some sort of force balance 
must hold between them. One can ask what are the different interactions which contribute to
this force balance?  In order to answer this question,
we will look at the forces between two widely separated IWP solitons in the limit
that one of the solitons is much lighter than the other and may therefore be 
treated as a test-object, or probe, moving in the background spacetime fields of the heavier soliton.

The force balance which operates in the MP case, when the metric function $f$ is real, is well
known.  It is simply the cancellation between the Newtonian gravitational attraction
and the Coulombic electrostatic repulsion for two solitons, each of which has charge equal to its mass. 
That this cancellation continues to hold exactly in general relativity, beyond the
Newtonian limit, is quite remarkable.

Now, consider taking the parameters $M_i$ to be complex, while keeping the
position vectors $\vec x_i$ real. In order to analyse the force balance in this case, 
we would need the equation of motion for a test particle with NUT charge. Some time ago,
it was conjectured that NUT charge may play a dual role to mass, in the 
same way that electric and magnetic charges are dual within Maxwell theory \nutdual\
(see also \zee ), with mass and NUT charge satisfying a Dirac-like quantization
condition. In this context, an equation of motion for a test particle carrying NUT
charge (but no mass) was suggested in \dowker , which is simply the geodesic equation
in a dual metric, defined to have Riemann curvature dual to that of the original
background metric.  This seems to be a good starting point for us. However, we would also need to include
ordinary mass for the test particle, as well as, electric and
magnetic charges. It seems likely that an ansatz for such an equation of motion could 
be formulated along the lines of \dowker . Whether, or not, this equation produced a
force balance for IWP solitons could then be regarded as a test of its validity.

We leave this direction to future work and concentrate instead on IWP solitons carrying angular momentum.
However, we do note here that the pairing in equation \nutrelation\ of mass with electric charge and NUT
charge with magnetic charge for IWP solitons fits in well with the conjectured duality of
mass and NUT charge described above. It also seems likely that a proper understanding of the motion of test
particles with NUT charge would contribute to a similar understanding of the motion of Kaluza-Klein monopoles
\rafael\gross\ and D6-branes (as described in \paul), which involve a Euclidean Taub-NUT space in their
construction.

We now turn to IWP solitons which carry angular momentum (but zero NUT charge). 
In this case, we require a spinning probe in the background of a single IWP soliton 
to carry out a force balance
analysis. Fortunately, the equation of motion
for a  spinning test particle has been investigated at great length in the literature. 
It was first derived in 1951 by Papapetrou 
\papapetrou\ by starting with finite
sized objects carrying angular momentum and taking a test particle limit. This work was extended by Dixon \dixon\ 
to include electromagnetic interactions giving the equation of motion
\eqn\spinning{\eqalign{
u^a\nabla_ap^b  &= qF^b{}_c u^c  -\half \left(  R^b{}_{cde}u^c +{g\over 2} {q\over
m} \nabla ^b F_{de} \right)   S^{de}\cr
& \equiv  F^b_{maxwell} + F^b_{spin}  +F^b_{dipole}\cr }}
Here $p^a$ and $u^a$ are the four-momentum and $4$-velocity of
a test-particle\foot{Far from the background soliton one has the usual relation
$p^a =mu^a$. However, in general, $p^a$ and $u^a$ are not collinear (see 
\wald\ for a discussion of this point).} which has mass $m$, charge $q$, gyromagnetic
ratio $g$ and angular momentum tensor $S^{ab}$. The angular momentum tensor  is related
to an angular momentum vector by $S_{ab}=\epsilon_{abcd}u^c S^d$. The term
$F_{maxwell}$ in \spinning\ gives the usual electromagnetic force. $F_{spin}$ and
$F_{dipole}$, we will see, give gravitational and electromagnetic spin-spin forces.

Wald \wald\ has computed $F_{spin}$ in \spinning\ for a test particle in the background
of a Kerr black hole ({\it i.e.} charge zero) with angular momentum $\vec{J}$.
In the limit of large separation, there is a spin-spin force on the test particle given by
\eqn\spinforce{
\vec{F}_{spin}=-\vec\nabla\left\{ {-\vS\cdot\vJ +3(\vS\cdot\hat{r})(\vJ\cdot\hat{r})\over
r^3}\right\}.}
Wald \wald\ notes that up to an overall sign change, this gravitational
spin-spin force has the same form as the familiar magnetic dipole-dipole force
from basic electrodynamics, if the spin vectors are replaced by magnetic moment vectors.

The spacetime fields for a single IWP soliton with vanishing NUT charge
are simply those of a Kerr-Newman spacetime with $Q=M$. It is easily checked that Wald's 
result \spinforce\ for $F_{spin}$ is
unchanged by the addition of charge for the background spacetime. 
For a charged test particle in the Kerr-Newman spacetime 
$F_{dipole}$ in equation \spinning\  will also be nonzero. 
To calculate this we plug into \spinning\
the far field limit of the spatial components of the gauge potential in Kerr-Newman
\eqn\potential{A_i\simeq {Q J_{ik} x^k \over M r^3 } .}
We then find that the magnetic spin-spin force, as expected from Wald's observation,
combines simply with the gravitational spin-spin force to give
\eqn\forcebal{
\vec F_{spin}  +\vec F_{dipole} =- \left(1-{gqQ\over 2mM}\right) 
\vec\nabla \left\{ {r^2\delta ^k _m -3 x^k x^m\over r^5} \right\}
J_{kn}S^{mn} .
}
One sees immediately that the sum of these forces vanishes, if
both the test particle and the black hole have charges equal to their masses,
$Q=M$ and $q=m$, and the test particle has gyromagnetic ratio $g=2$.
These are precisely the conditions for both the background and the probe to be IWP solitons. 
Note that the gyromagnetic 
ratio $g=2$ for the background is already built into the gauge potential \potential\
for the Kerr-Newman background. 

For the static probe, $F_{maxwell}$  in \spinning\ 
contributes a Coulombic force, which is balanced by the Newtonian gravitational force
contained in the the Christoffel symbol terms on the left hand side as in the MP case.  Putting this
together with the cancellation of the gauge and gravitational spin-spin forces in \forcebal, we then have 
a complete understanding of the force balance between spinning IWP solitons.

\newsec{Force Balance for BPS Superparticles}
We now turn to our second example. In reference \ae\ Aichelberg and Embacher 
constructed a class of multi-soliton solutions,  which they called ``superpartner'' spacetimes, in
the following way.  Start with the MP spacetimes. 
Since these break half the supersymmetries of
$N=2$ supergravity, one can generate new solutions by acting with the broken supersymmetry
generators. The resulting superpartner spacetimes have nontrivial gravitino fields and carry
nonzero, quantum mechanical, spin angular momenta, filling out a BPS multiplet of spin states\foot{
As discussed in \bktw\ the angular momentum is carried by the quantized states of the gravitino
zero-modes}.

In the further series of papers \aeone -\aefour\ Aichelburg and Embacher went on
to study the motion of a $\kappa$-symmetric superparticle probe in the background of a 
superpartner soliton. The superparticle represents the test particle limit of a second
superpartner soliton. This is precisely the setup we need to study the force balance
between superpartner solitons
We now show that it is a simple
application of the results of \aeone -\aefour\ that, 
in the limit of large separation between a superparticle and some number of 
superpartner solitons, a cancellation between gauge and gravitational spin-spin forces takes place
which allows static configurations to exist.

A lengthy calculation in \aethree\ 
yields the equation of motion for the superparticle, expanded out to
quadratic order in its fermionic superspace coordinates $\theta^k$. Aichelberg and Embacher identify
within this result a gravitational spin-spin interaction term, in which the
classical angular momentum of the probe in Papepetrou's equation of 
motion \papapetrou\ is replaced by the quantum mechanical spin operator of the
superparticle. Here we additionally note that the superparticle equation of motion in \aethree\
also  contains a magnetic dipole-dipole force term, which cancels with the gravitational spin-spin force
as part of the force balance between superpartners.
From equation (6.7) of \aeone\
we have\foot{Additonal force terms, besides those we display
here also appear in the equation of motion of the superparticle in \aeone . That all of these terms give
zero net interaction between superpartners is noted at the level of the superparticle lagrangian in \aefour\ 
without a specific discussion of the nature of the cancelling forces.}
\eqn\supforce{
F^{spin}_{\mu}+ F^{dipole}_{\mu}  ={1\over 4} \bar{\pi}^j u^{\nu}
\left[ \delta ^{jk} R_{\mu\nu}^{~~mn}
\gamma _{mn}  -{iq\over m} \epsilon ^{jk}\gamma ^{mn} (\partial _{\mu} F_{mn} \gamma _{\nu}
- \partial _{\nu} F_{mn} \gamma _{\mu} ) \right] \theta ^k
}
Here $\theta ^j$ , $j=1,2$ are the two (Majorana) fermionic coordinates of the superparticle, 
$\pi^j$ are their conjugate momenta, 
$\gamma_{\mu}$ are four dimensional gamma matrices, and $\gamma _{mn}$ is
the antisymmetric product of Dirac matrices. Latin letters are frame indices and greek
letters are spacetime coordinate indices. We refer the reader to \aeone\ for detailed
definitions and conventions\foot{We have substituted in the relation between the superparticle 
supercharge $Q^k$ and the fermionic coordinates $\theta^k$, $Q^k=\pi^k +O( \theta ^2 )$ into equation (6.7) of \ae\ to
obtain \supforce .}.

The spin operator of the superparticle $S^{op}_{mn}$, to order $\theta^2$, is  given by \ae\ 
\eqn\supspin{
S^{op}_{mn}= - \bar{\pi}^j \gamma_{mn} \theta ^j .}
With this identification, the first term in equation \supforce\ 
has exactly the same form as the classical gravitational
spin-spin force $F_{spin}$ in equation \spinning\ above. To study how the force balance occurs in the
superpartner spacetimes, consider a static probe, as before, with four velocity  $u^m =(1,0,0,0)$ and
look at the gauge field strength terms in \supforce.
The spinor coordinates $\theta^k$ in \aeone, for this four velocity, satisfy the gauge condition 
$(1+\gamma _0) (\theta ^1 +i\theta ^2 )= 0$ and using the fact that $\gamma _0$ is pure imaginary in \aeone  ,
it then follows that $\bar{\pi}^j \epsilon _{jk} \gamma _0 \theta ^k =i\bar{\pi}^j \delta _{jk} \theta ^k$.
Substituting this and the static property of the background $\partial _0 F_{mn} =0$ into \supforce, 
we finally find that the sum of the
gravitational and magnetic spin-spin forces for the superparticle in equation 
\supforce\ reduces {\it identically}
to the expression \forcebal\ for the classical spinning particle, with the spin tensor of
the classical particle replaced by the spin operator $S^{op}_{mn}$ of the superparticle.
The gyromagnetic ratio for the superparticle automatically comes out to be $g=2$.

The asymptotic forms of the metric and gauge potential of the
superpartner spacetimes of \ae\ are the same as that of Kerr-Newman with $Q=M$.  Therefore, the
evaluation of the Riemann tensor and electromagnetic field strength in \forcebal\  is as before.
The gravitational
spin-spin and magnetic dipole-dipole forces then balance for the superparticle in these spacetimes
in the same way they do for the classical spinning particle in the IWP spacetimes. 
We note that the superparticle spin in \supspin\ is operator valued.
To get actual numerical values for the individual forces in equation \supforce , 
one must evaluate the force in a particular spin state. 
This and related issues are discussed in \aefour\bktw.

\bigskip
{\bf Acknowledgements: } 
We would like to thank Vijay Balasubramanian for helpful discussions on these topics.
This work was supported in part by NSF grant NSF-THY-8714-684-A01.

\listrefs
\end